\input{aipcheck}

\documentclass[
    ,final            
  ]
  {aipproc}

\layoutstyle{8x11single}

\def\beq{\begin{eqnarray}}
\def\eeq{\end{eqnarray}}

\begin{document}

\title{Cross section analyses in MiniBooNE and SciBooNE experiments}


\author{Teppei Katori for the MiniBooNE and SciBooNE collaborations}{
   address={Massachusetts Institute of Technology, Cambridge, MA}
}

\begin{abstract}
The MiniBooNE experiment (2002-2012) and the SciBooNE experiment (2007-2008) 
are modern high statistics neutrino experiments, 
and they developed many new ideas in neutrino cross section analyses. 
In this note, I discuss selected topics of these analyses. 
\end{abstract}

\maketitle


\section{The MiniBooNE experiment}

The MiniBooNE experiment uses the Booster Neutrino Beamline (BNB)~\cite{MB_flux} 
and the MiniBooNE detector~\cite{MB_dtec}. 
The BNB creates $<E_\nu>\sim$700 (600)~MeV muon neutrino (anti-neutrino) beams. 
The neutrinos (anti-neutrinos) travel $\sim$520~m before observed by the MiniBooNE Cherenkov detector 
filled with mineral oil (CH$_2$), through productions of charged particles.  

\subsection{Signal definition}

How to define a signal event is important for any cross section measurement. 
We encountered a problem when the charged-current 1$\pi^+$ production (CC1$\pi^+$) 
to charged-current quasielastic (CCQE) cross section ratio was studied~\cite{MB_ccpipratio}. 
The final state interactions (FSIs) affect the pion observation, through 
pion absorption, charge exchange, pion production, and re-scattering. 
All of them have large errors.  
How to remove these FSI effects to measure genuine pion kinematics 
from the neutrino interaction vertices?
The answer we came was not to remove FSI effects, 
but define our signals differently. 
Since the FSIs are not understood well, 
any corrections on FSIs introduce extra biases in the data and make the data model dependent. 
Instead, we define the signals from the final state particles in the detector. 
In the case of the CC1$\pi^+$ interaction, a signal event is defined by 
``1$\mu$ and 1$\pi^+$ in the final state''. 

\begin{figure}[b]
\includegraphics[height=.15\textheight]{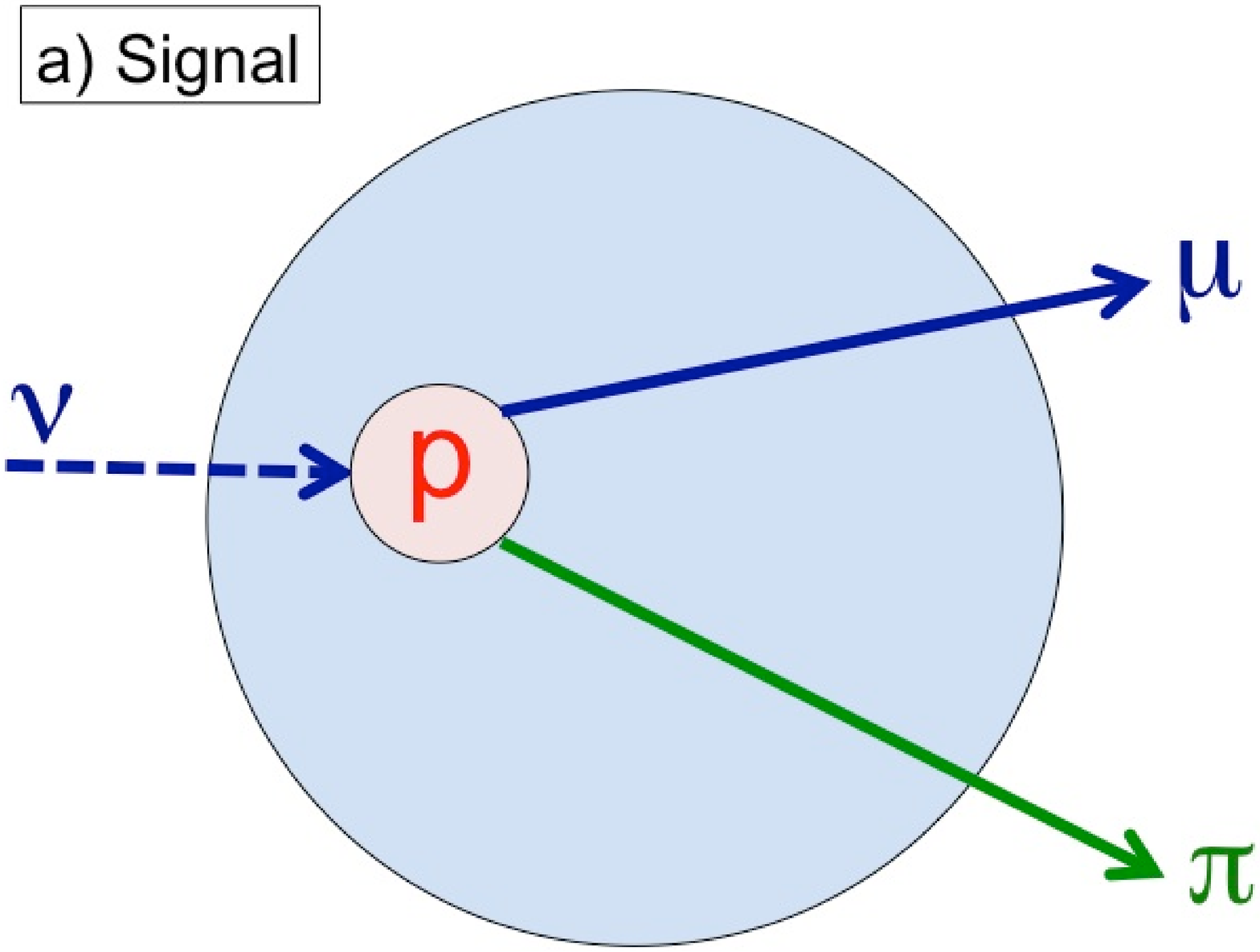}
\includegraphics[height=.15\textheight]{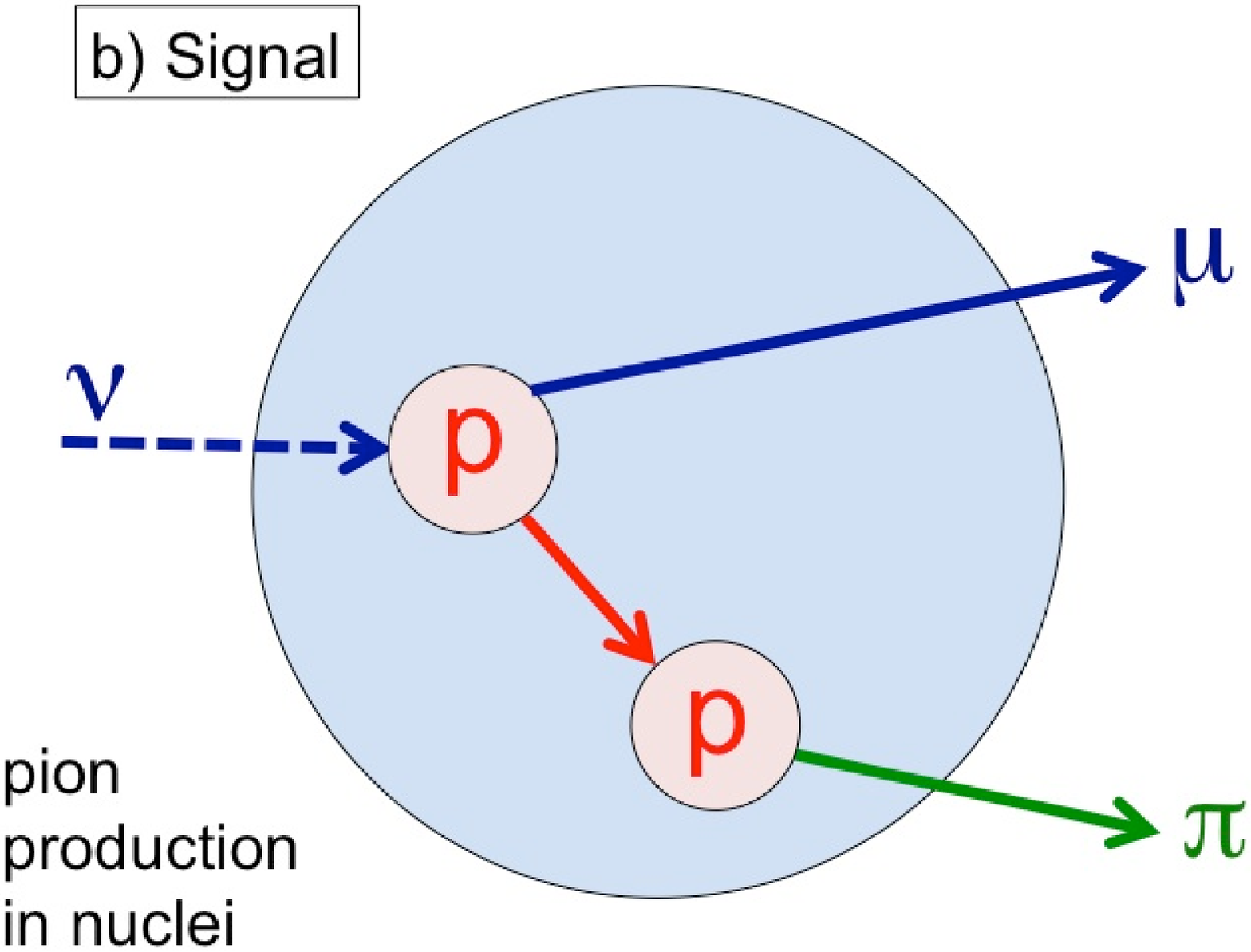}
\includegraphics[height=.15\textheight]{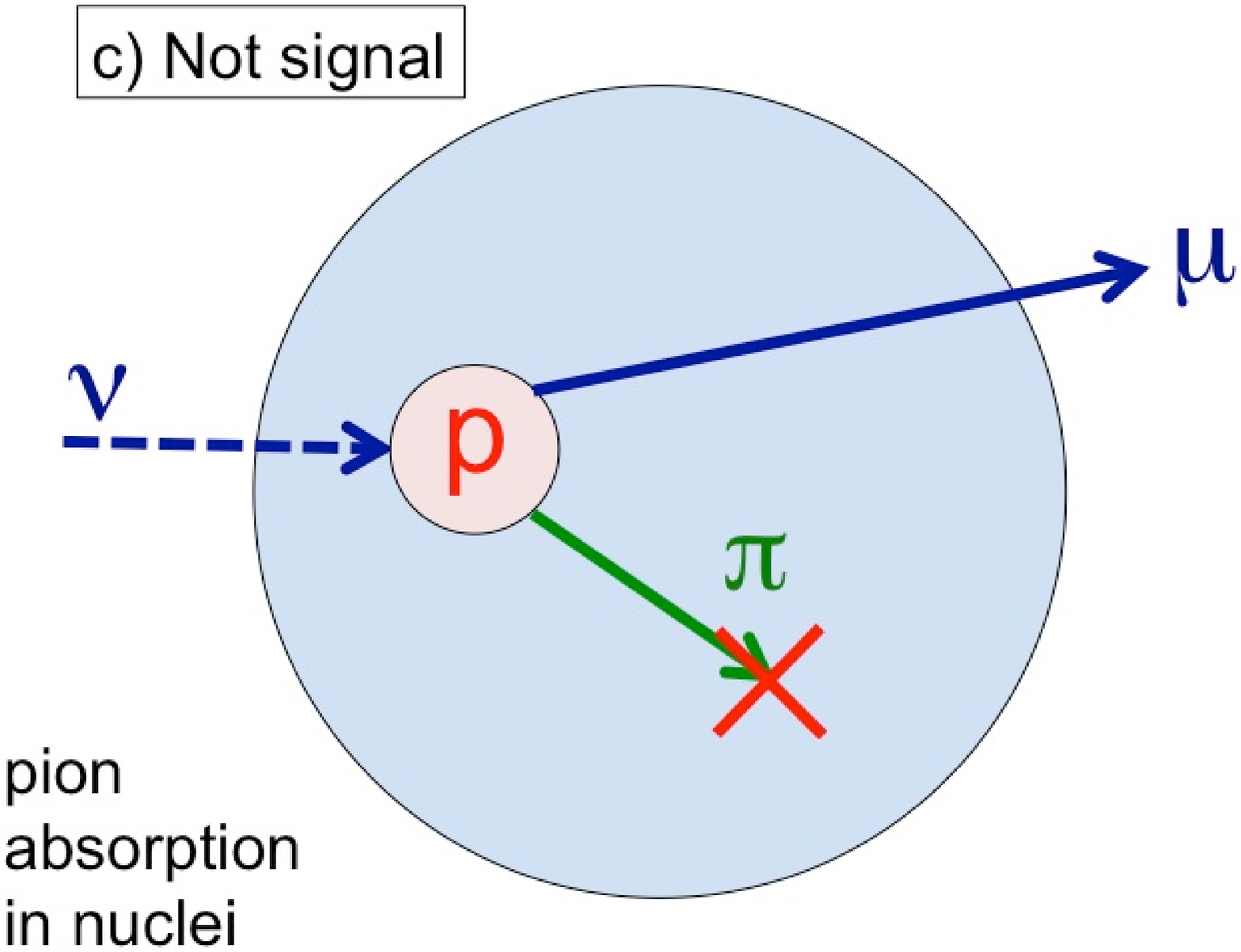}
\includegraphics[height=.15\textheight]{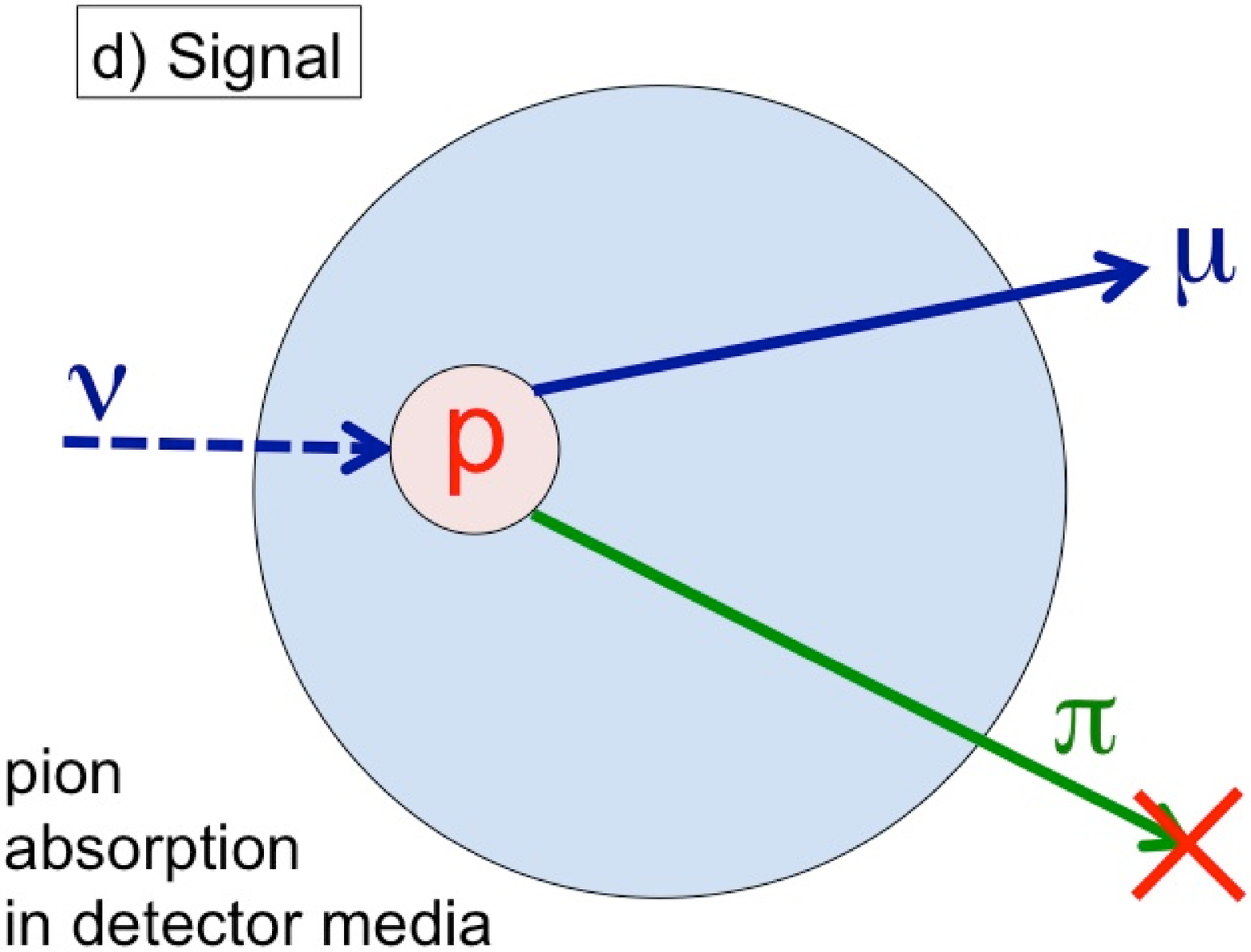}
\caption{\label{fig:sigdef}
Signal definition of pion production measurement. 
(a), (b), and (d) are classified to be ``signal'', and (c) is ``not signal''. 
Notice both (c) and (d) are pion absorptions, but (b) is in the target nuclei, 
and (d) is in the detector media.}
\end{figure}

Figure~\ref{fig:sigdef} shows this situation. 
Fig.~\ref{fig:sigdef}a shows a CC process accompanied with one pion production, 
where both the muon and the pion leave the nuclear target and are observed successfully. 

However, some final-state pions may also be created not by the primary neutrino interaction, 
but by FSI processes, such as the hadronic re-interaction (Fig.~\ref{fig:sigdef}b). 
Since our detector cannot distinguish such a pion from those pions created from the primary interaction, 
such event must also be included in what we define as "signal".
The data is a sum of both, which help theorists to study FSIs. 

Figure~\ref{fig:sigdef}c shows an opposite case, 
where pions made by the primary neutrino interaction fails 
to leave the target nuclei, 
hence are not observed. 
Traditionally, experiments estimate how many pions disappeared due to FSI, 
and apply corrections. 
However, such corrections are model dependent and should be avoided. 
By our signal definition, this kind of event is classified as ``not signal''.

The last example (Fig.~\ref{fig:sigdef}d) is a cumbersome situation. 
Pions can be absorbed during their propagation in the detector medium. 
Since this is a detector dependent effect, we need to correct based on our simulation. 
This introduces an error, and for example, 
such detector dependent nuclear error dominates 
the MiniBooNE CC 1$\pi^{\circ}$ production (CC1$\pi^\circ$) measurement~\cite{MB_ccpi0}.

\begin{figure}[tb]
\includegraphics[height=.25\textheight]{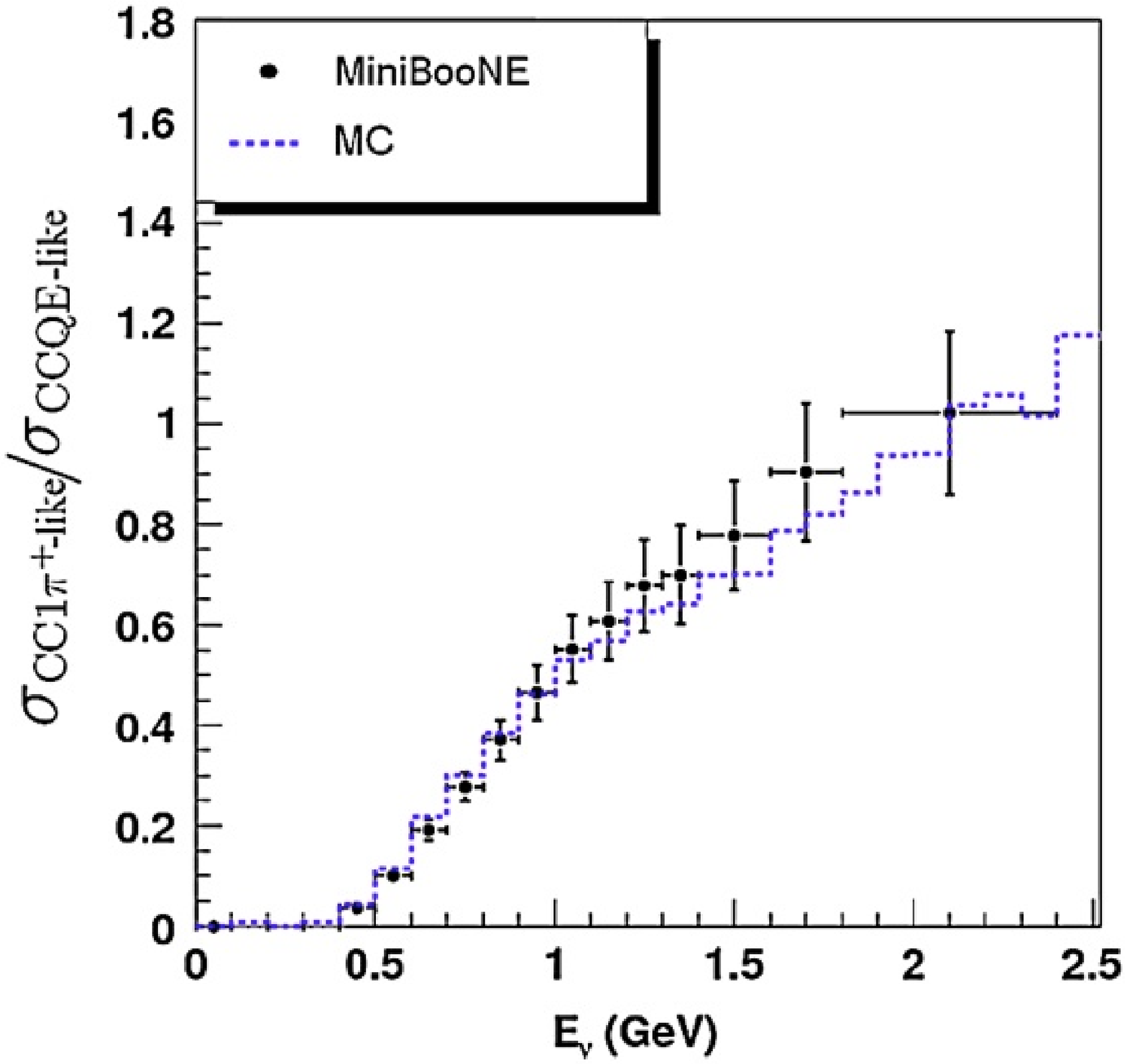}
\includegraphics[height=.25\textheight]{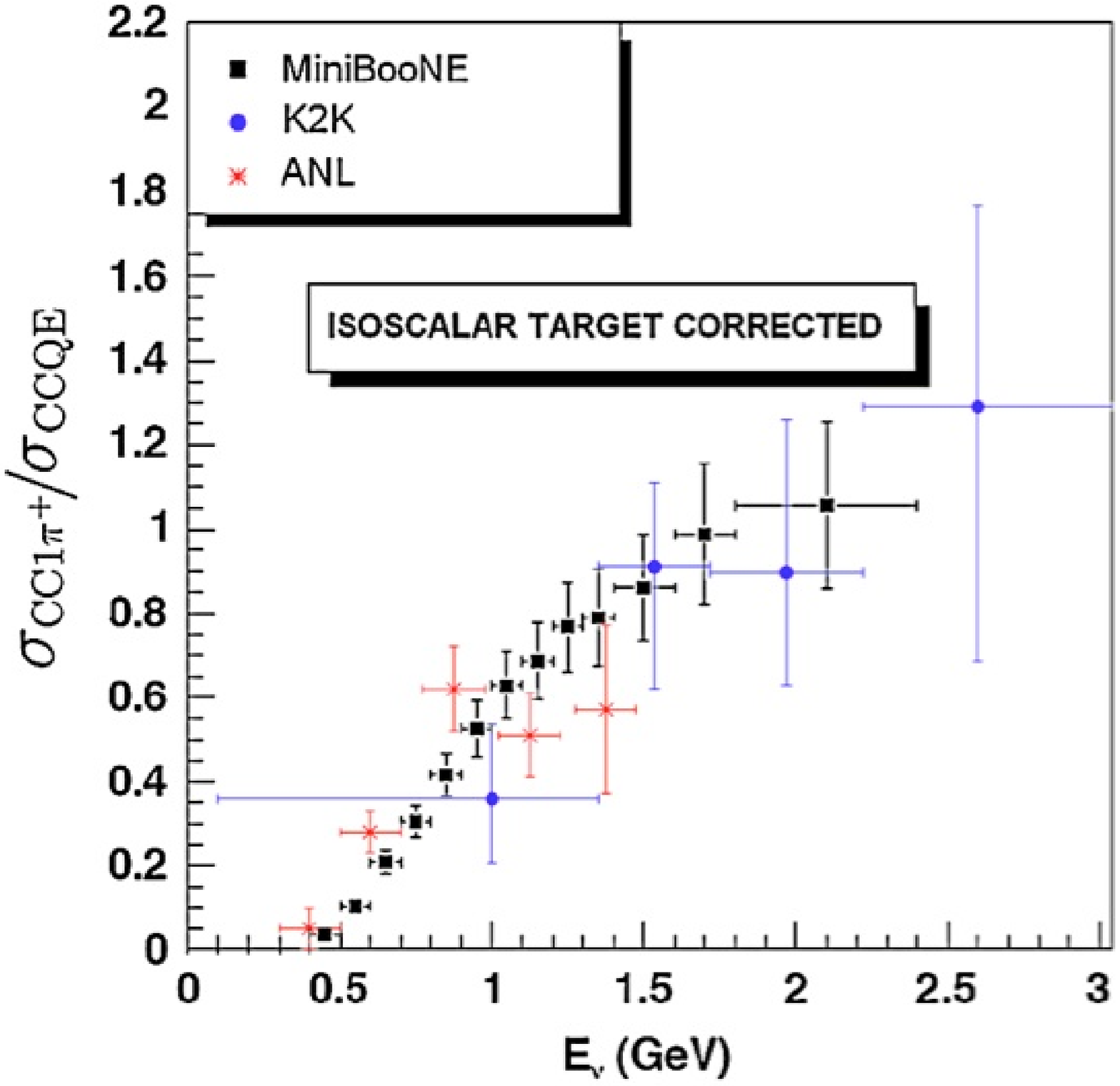}
\caption{\label{fig:ccpipratio}
MiniBooNE CC1$\pi^+$/CCQE total cross section ratio~\cite{MB_ccpipratio}. 
In the left plot, there is no correction applied about FSI in the target nuclei, 
and hence, reactions are called ``CC1$\pi^+$-like'' and ``CCQE-like''. 
In the right plot, the simulation dependent correction is applied.
}
\end{figure}

The signal definition affects how the final data appear. 
Fig.~\ref{fig:ccpipratio} shows the CC$1\pi^+$/CCQE total cross section ratio 
taken from Ref.~\cite{MB_ccpipratio}. 
These 2 total cross section ratio plots are based on different signal definitions. 
On the left, signal is defined from observables (Fig.~\ref{fig:sigdef}), 
instead of primary interactions, 
and hence the measured total cross section ratio is called 
``CC1$\pi$-like'' to ``CCQE-like'' ratio. 
This result is independent of any nuclear models. 
The price to pay is, theorists can compare their models with our data 
only if they have ways to simulate FSI effects. 
Otherwise, theorists are recommended to use the right plot. 
Here, simulation dependent corrections of the target nuclei FSI are applied, 
therefore the ratio can be interpreted as a primary interaction cross section ratio. 
Historically only such measurements are performed~\cite{ANL_ccpipratio,K2K_ccpipratio}. 
The price to pay is, the resulting ratio is nuclear model dependent, 
and the error bars are inflated to take into account this model dependency. 

\subsection{Data driven correction}

Measured interactions always include background events. 
And these background events need to be removed, 
however, this depends on the predictions from the simulations. 
To avoid such model dependency, 
MiniBooNE actively uses the data driven corrections to the background events, 
except for the CC1$\pi^+$ cross section measurement 
where the signal purity is $\sim$90\%~\cite{MB_ccpip}. 

\subsubsection{CCQE cross section measurement}
In the CCQE cross section measurement~\cite{MB_ccqe}, CC1$\pi^+$ events make the biggest background. 
If a CC1$\pi^+$ event loses a pion through FSI, 
its final state particles are identical with those of CCQE events. 
Simultaneous measurements of the CCQE and CC1$\pi^+$ candidate sample are performed, 
and information from the CC1$\pi^+$ candidate sample allows 
the background in the CCQE candidate sample 
to be corrected as a function of reconstructed 4-momentum transfer ($Q^2$). 
This corrected background is subtracted from the CCQE sample 
to measure various cross sections. 
Note this correction is valid only within the precision of FSI. 
To allow theorists to study FSIs, 
MiniBooNE also published the subtracted background distributions~\cite{MB_ccqe}.  

\begin{figure}[tb]
\includegraphics[height=.2\textheight]{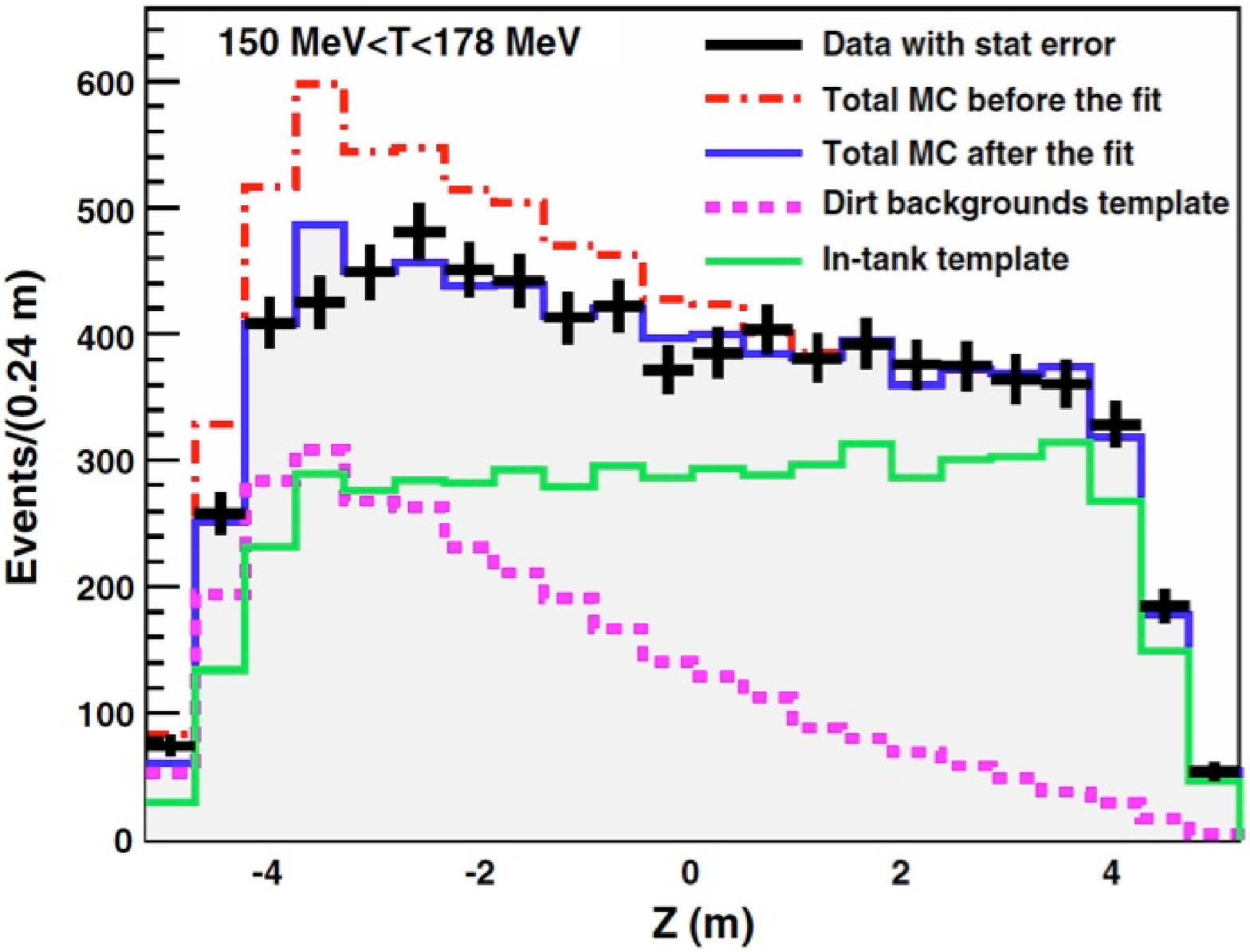}
\includegraphics[height=.2\textheight]{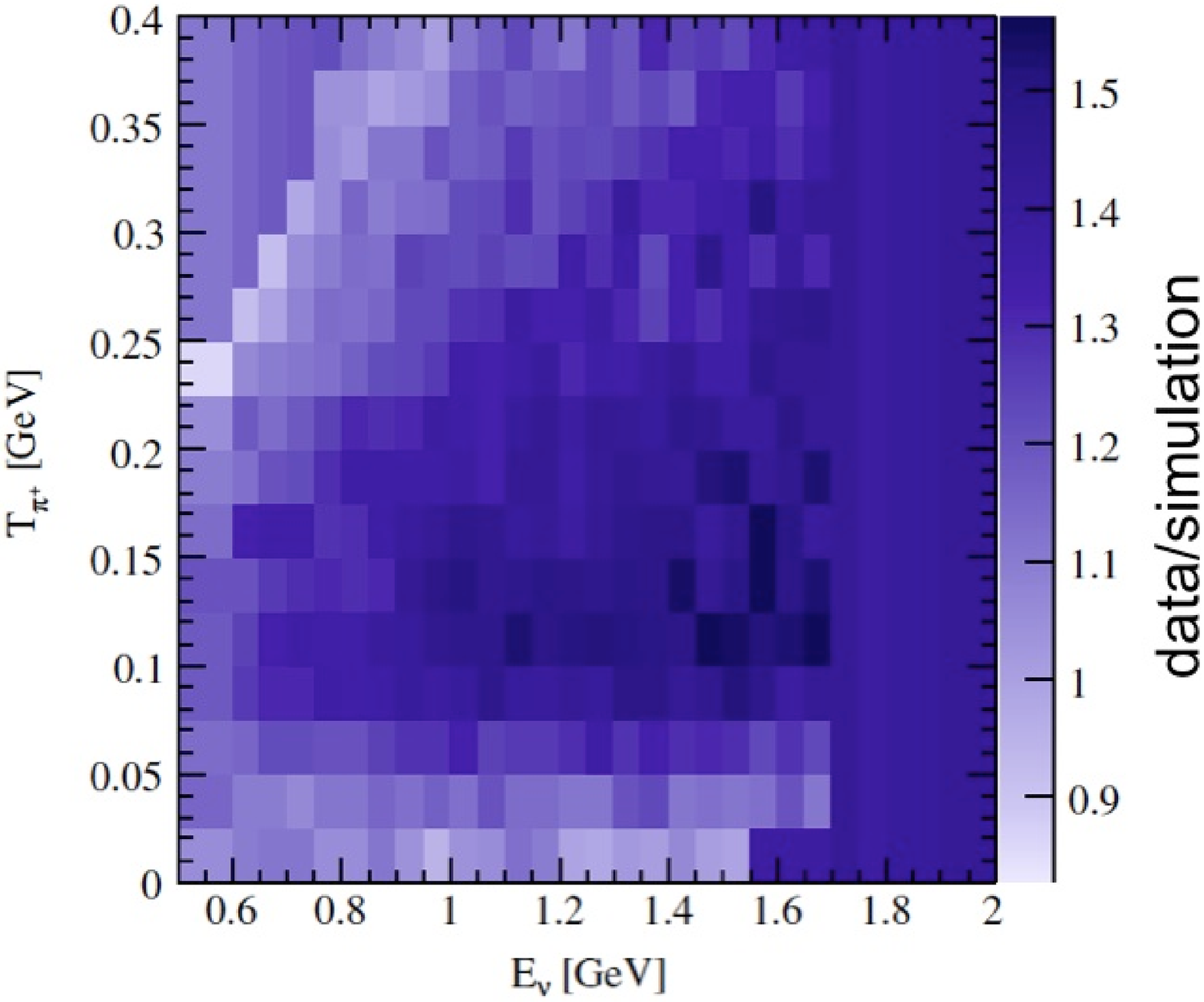}
\includegraphics[height=.2\textheight]{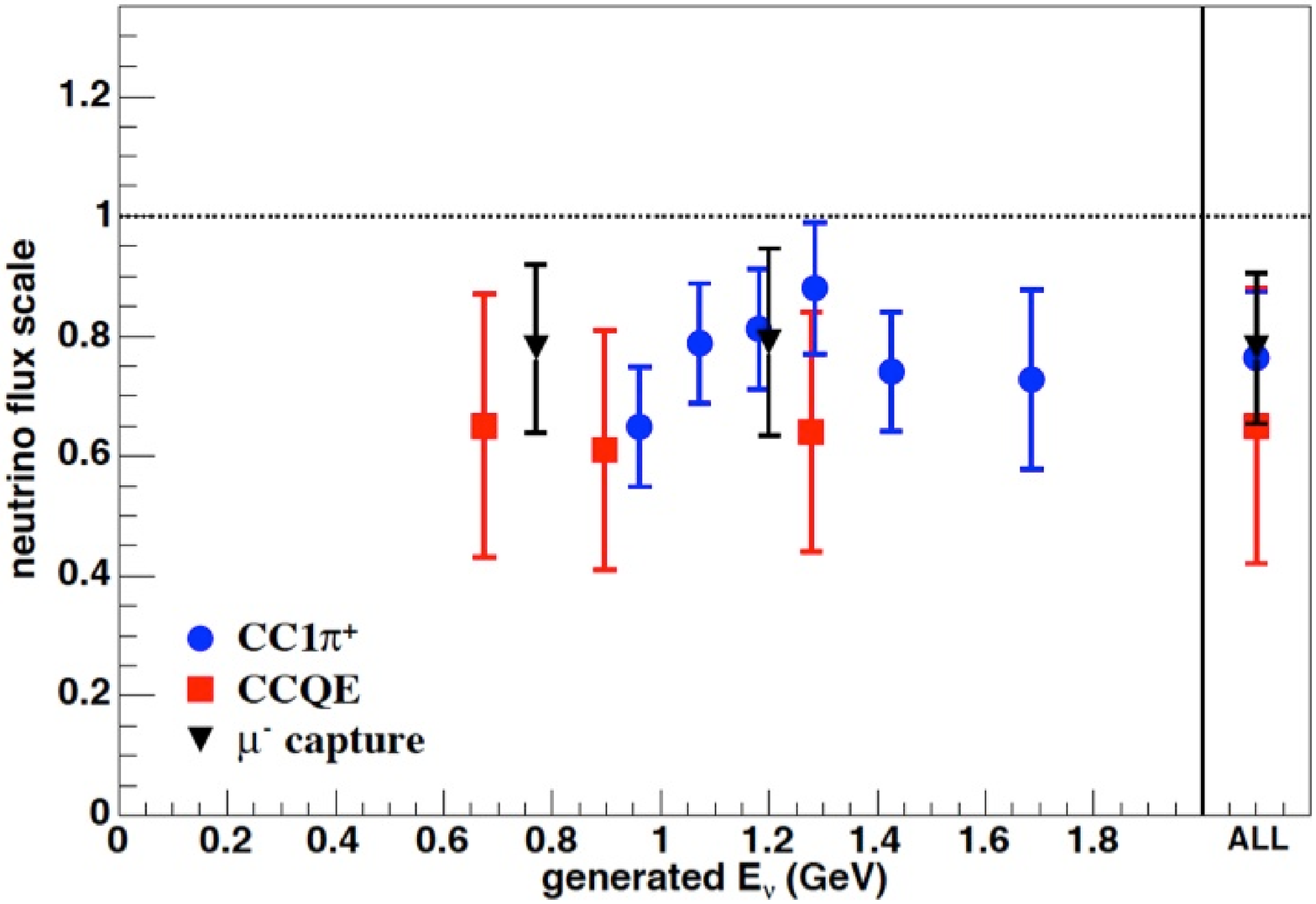}
\caption{\label{fig:ddc}
Examples of MiniBooNE data driven corrections. 
In the left plot, dirt event enhanced NCEL sample is shown with various components from the simulation~\cite{MB_ncel}. 
The template fit corrects the scaling factor of dirt events.
In the middle plot, the ratio of measured CC1$\pi^+$ to simulated CC1$\pi^+$ double differential cross section 
is shown~\cite{MB_ccpi0}. 
This is used to correct the CC1$\pi^+$ background for CC1$\pi^\circ$ cross section measurement.
In the right plot, flux scale factors of neutrino component 
in the anti-neutrino beam are shown as a function of $E_\nu$~\cite{MB_anticcqe,MB_ws}. 
There are 3 independent measurements and they are consistently lower than 1 (default simulation). 
}
\end{figure}

\subsubsection{Neutral current elastic (NCEL) cross section measurement}
In the NCEL cross section measurement~\cite{MB_ncel}, 
backgrounds created outside of the detector (dirt events) are significant. 
Fortunately, the MiniBooNE detector is big enough to see 
the spatial profile of dirt events in the detector.  
Figure~\ref{fig:ddc}, left, shows the dirt event enhanced NCEL sample 
as a function of the vertex location in Z (beam direction)~\cite{MB_ncel}. 
The template fits can find the normalization factor of dirt events, 
as a function of measured total nucleon kinetic energy. 
After scaling the dirt events, data and simulation agree well.  
This technique to correct dirt events is also used for the oscillation analysis~\cite{MB_osc}. 

\subsubsection{CC1$\pi^\circ$ cross section measurement}
In the CC1$\pi^\circ$ cross section measurement~\cite{MB_ccpi0}, 
CC1$\pi^+$ events make large backgrounds. 
Unlike the CCQE cross section measurement where 
the CC1$\pi^+$ background is corrected as a function of reconstructed $Q^2$, 
here the correction is based on the double differential cross section of 
pion kinetic energy and neutrino energy ($E_\nu$). 
Figure~\ref{fig:ddc}, middle, shows the data to simulation ratio~\cite{MB_ccpi0}, 
where data is the measured CC1$\pi^+$ double differential cross section~\cite{MB_ccpip}. 
This ratio is applied as a correction to the simulated CC1$\pi^+$ background. 

\subsubsection{Neutral current 1 $\pi^\circ$ (NC1$\pi^\circ$) cross section measurement} 
In the case of the NC1$\pi^\circ$ cross section measurement~\cite{MB_ncpi0}, 
no data driven correction is applied. 
However, the measured rate of NC1$\pi^\circ$ is used to correct 
the background distribution of the oscillation candidate events~\cite{MB_osc}. 
Here, $\pi^\circ$ background in the oscillation sample 
is corrected as a function of $\pi^\circ$ momentum~\cite{MB_ncpi0ratio}. 
This measurement not only corrects the rate, 
but also constrains the $\pi^\circ$ rate error, 
because the cross section error of NC1$\pi^\circ$ is $\sim$25\% from the simulation, 
whereas the measured NC1$\pi^\circ$ rate has $\sim$5\% error. 

\subsubsection{Anti muon-neutrino CCQE cross section measurement}
The $\bar\nu_\mu$CCQE cross section measurement~\cite{MB_anticcqe} 
has a complicated structure, 
because of the presence of neutrino contamination in the anti-neutrino beam. 
This is measured, and corrected by using 3 independent techniques~\cite{MB_anticcqe,MB_ws}. 
Fig.~\ref{fig:ddc}, right, shows the result. 

The CC$1\pi^-$ production from anti-neutrino interactions 
make large backgrounds for the $\bar\nu_\mu$CCQE cross section measurement. 
The simultaneous fit technique used in the $\nu_\mu$CCQE measurement 
cannot be applied since the CC$1\pi^-$ final state contains same final state particles with $\bar\nu_\mu$CCQE 
($\pi^-$ is $\sim$100\% nuclear captured). 
Therefore the CC1$\pi^-$ distribution is corrected by assuming the same kinematic distribution 
as the CC$1\pi^+$ distribution in neutrino mode. 

Finally the $\bar\nu_\mu$CCQE cross section model is corrected based on 
the relativistic Fermi gas model tuning~\cite{MB_ccqeprl} from the $\nu_\mu$CCQE measurement~\cite{MB_ccqe}.  
Note that the absolute cross section measurement does not depend on the cross section model of the signal channel 
(in this case, $\bar\nu_\mu$CCQE model), except in unfolding. 

The $\bar\nu_\mu$NCEL cross section measurement~\cite{MB_antincel} is underway.  
With this result, MiniBooNE will complete all 4 quasielastic and elastic scattering cross section measurements. 
The difficulty is, our NCEL cross section measurements are not on a proton or a neutron target, 
but the combination of both. This makes it difficult to apply interesting ideas to access to 
nucleon parameters~\cite{Alberico,Jachowicz}. 
Theorists are encouraged to invent new ways to utilize MiniBooNE CCQE and NCEL data!

\subsection{Background removing process}

\begin{figure}[tb]
\includegraphics[height=.2\textheight]{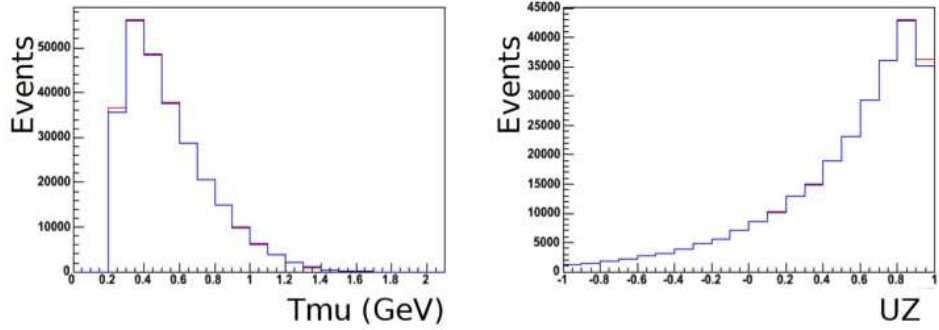}
\caption{\label{fig:bsub}
The MiniBooNE CCQE signal sample based on different background removing methods~\cite{Teppei}. 
The red lines are for background subtraction method, and the blue lines are for signal fraction method. 
}
\end{figure}

Background events need to be removed from the data, 
but depending on the knowledge of the estimated background, 
there are several ways to remove the background. 

If the knowledge of the background events includes the absolute scale 
(for example, backgrounds are measured {\it in situ}), 
the background subtraction method is applied ($d_i-b_i$). 
In this case, signal events in $i$th bin is 
simply a difference of $i$th bin of data minus $i$th bin of background.   
This is a simple and preferred method, 
because it does not depend on the simulation of the signal events. 
The drawback is, you may get negative bins. 

If the knowledge of the background is at most a fraction of the total event, 
the signal fraction method is applied ($d_i\times\frac{s_i}{s_i+b_i}$). 
In this case, signal events in $i$th bin is 
data times predicted signal events ($s_i$) 
divided by predicted total events ($s_i+b_i$) of the simulation.   
The drawback is, the signal sample becomes signal model dependent. 

Figure~\ref{fig:bsub} shows the example of these two methods 
from the MiniBooNE $\nu_\mu$CCQE analysis~\cite{Teppei}.  
Since the data-simulation agreement is good in entire kinematic region, 
the difference of these two methods is small in this sample. 

\subsection{Unfolding}

\begin{figure}[tb]
\includegraphics[height=.22\textheight]{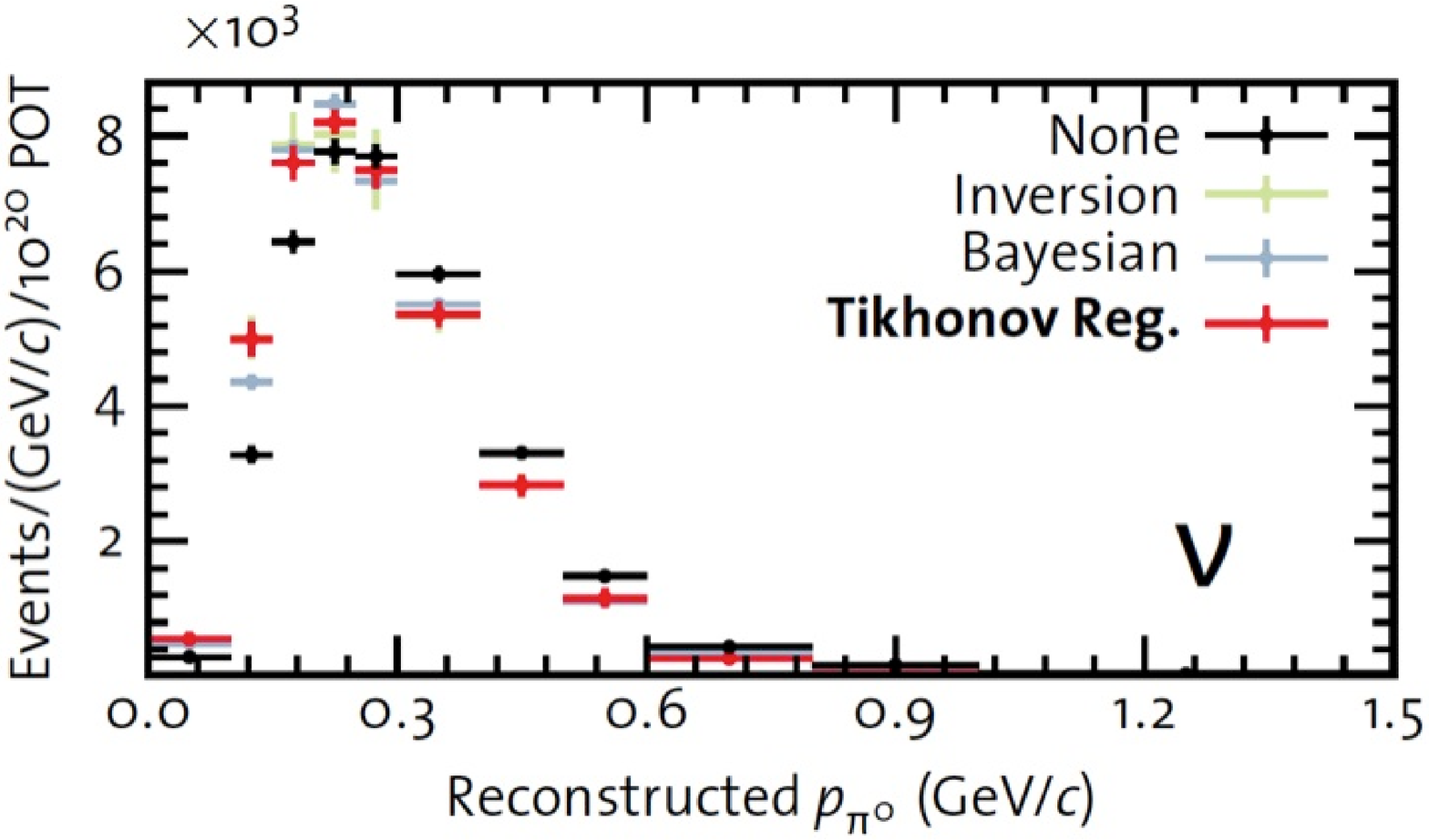}
\includegraphics[height=.22\textheight]{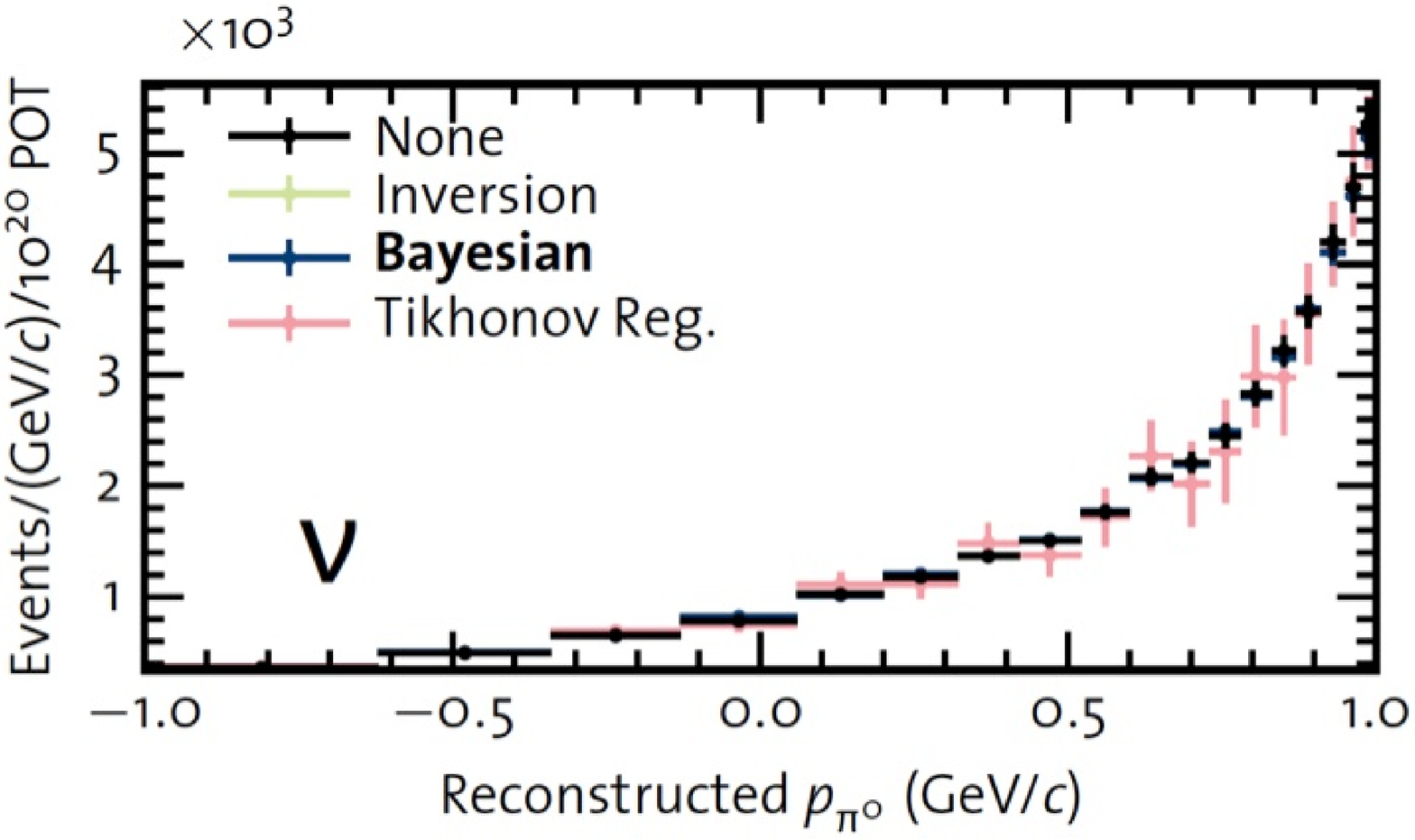}
\caption{\label{fig:unfold}
The $\pi^\circ$ kinematics unfolding from NC1$\pi^\circ$ cross section measurement~\cite{Colin}. 
Left plot shows $\pi^\circ$ momentum, and the right plots is for $\pi^\circ$ angular distribution.
}
\end{figure}

Unfolding is an important stage of the analysis, 
because measured kinematics are often smeared or distorted by detector effects. 
These detector effects need to be corrected by the data unfolding. 
Most of MiniBooNE cross section data are unfolded by the iterative Bayesian method~\cite{DAgostini},  
to avoid the fast oscillation problem which is often seen 
for histograms with many bins unfolded by the inverse matrix method~\cite{Denis}. 
The unfolding technique depends on many factors, 
and every single histogram needs to be assessed for the best unfolding scheme. 
Figure~\ref{fig:unfold} shows NC1$\pi^\circ$ kinematic distributions based on 
different unfolding techniques~\cite{Colin}. 
In this analysis, 
Four different unfolding techniques (one of four is no unfolding) are used to compare results. 
In the left, Tikhonov regularization~\cite{Tikhonov} is chosen to unfold $\pi^\circ$ momentum, 
but the same technique does not work for $\pi^\circ$ angular distribution (right), 
and the iterative Bayesian method is chosen for the main result. 

\section{The SciBooNE experiment}

The SciBooNE detector is an X-Y tracker at the BNB~\cite{MB_flux}. 
The detector consists of 3 parts, the Scintillation bar vertex detector 
(SciBar, consists of C$_8$H$_8$), 
electron catcher (EC), and muon range detector (MRD)~\cite{SB_ccpip}. 
Inclusive CC and CCQE total cross section results show similar excesses 
to the MiniBooNE cross section results~\cite{SB_ccincl,SB_ccqe}. 

\subsection{Event classification}

\begin{figure}[t]
\includegraphics[height=.20\textheight]{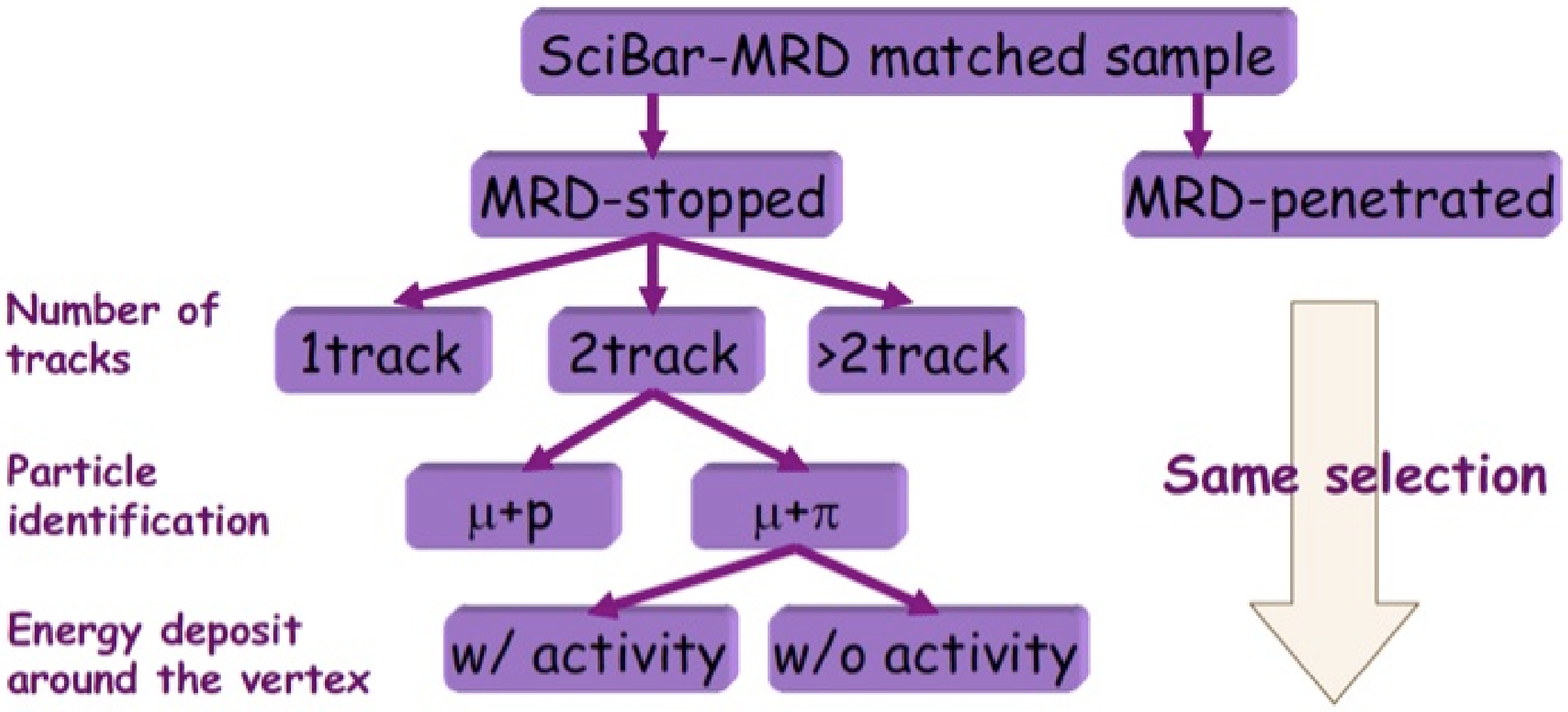}
\includegraphics[height=.20\textheight]{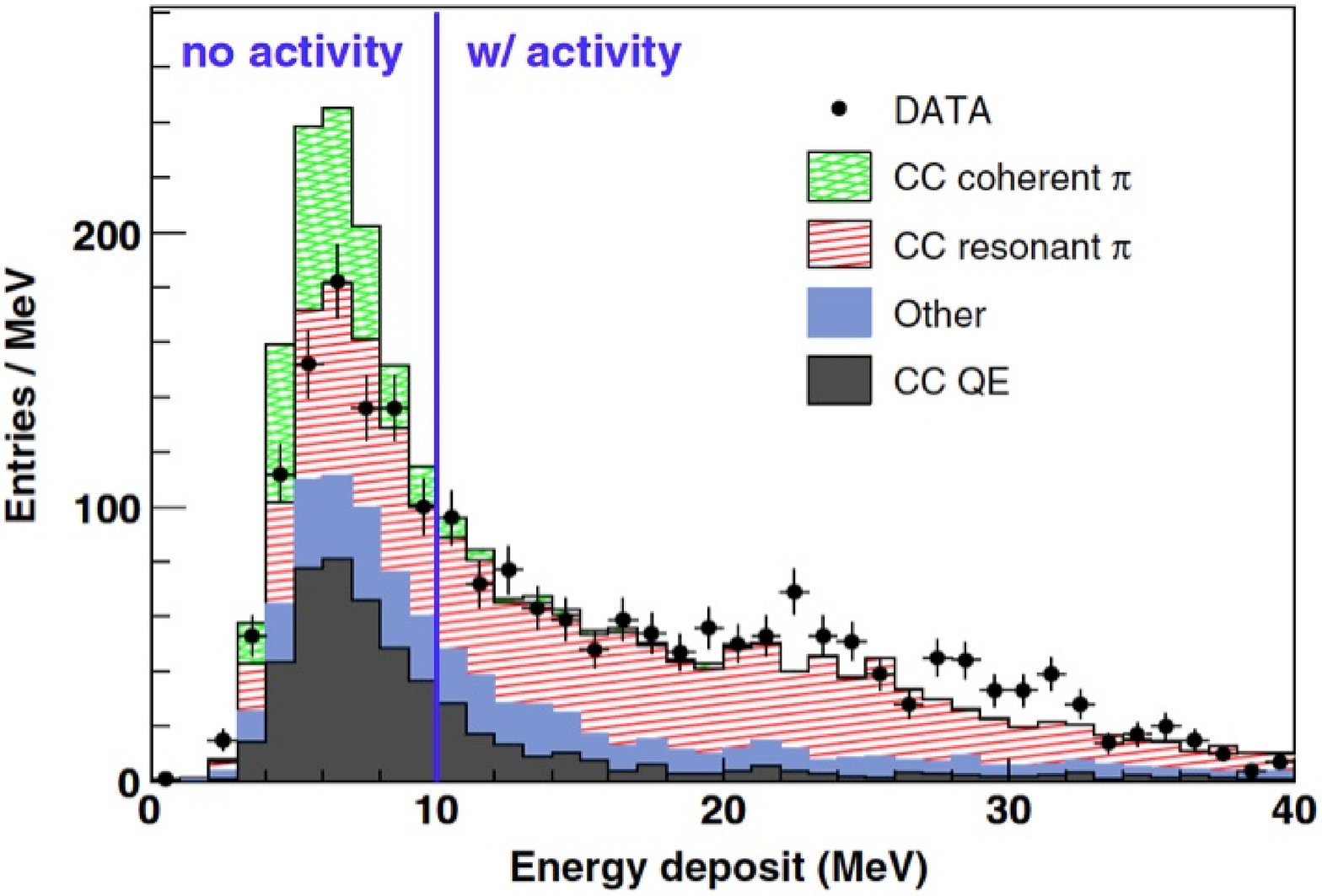}
\caption{\label{fig:class}
Flow chart of event classification for the CC1$\pi^+$ analysis (left). MRD stopped, 2 tracks, 
1 muon-like and 1 pion-like sample is shown on the right plot.
}
\end{figure}

SciBooNE can classify each event based on the topology. 
Figure~\ref{fig:class} left, shows the CC1$\pi^+$ analysis flow chart of event classification~\cite{bancho}. 
An event is classified depending on the type of tracks, number of tracks, and amount of vertex activity 
(energy deposit around the neutrino interaction vertex). 
Fig.~\ref{fig:class} right plot shows the event sample with 
``1 muon and 1 pion tracks'' in the final state, 
where the lower vertex activity region has more  
coherent CC1$\pi^+$ production according to the simulation. 
However, the data is consistent with the absence of coherent pions, 
as first observed by the K2K experiment~\cite{K2K_ccpip}.
The vertex activity is a powerful parameter, 
and similarly, 
NC1$\pi^\circ$ analysis confirmed the presence of 
coherent production by the same technique~\cite{SB_ncpi0}. 
This information is used for the latest T2K oscillation analysis~\cite{T2K_2013}.

The high resolution of the SciBar detector allows detailed study of each recorded track. 
The azimuthal angular distribution of pions 
may reveal further information of coherent pion production mechanism~\cite{SB_anticcpi}. 
Single proton momentum measurement from the NCEL scattering~\cite{SB_ncel} 
allows neutrino energy reconstruction without lepton kinematics. 
SciBar can also tag complicated topologies, such as CC1$\pi^\circ$~\cite{SB_ccpi0}. 
All these are possible due to the high resolution of the SciBar detector. 
When more tracks are measured in  more detail, 
an event can be classified in even smaller sub divisions. 
This eventually allows one to study the detailed structure of the FSIs. 
In this conference, ArgoNeuT showed an event-by-event counting of protons from CC interaction. 
Future high resolution experiments, such as MicroBooNE~\cite{georgia}, 
will allow further classification to understand more details of neutrino interactions.

\begin{theacknowledgments}
The author thanks the organizer for the invitation to the conference, 
and the hospitality during my stay in Rio de Janeiro. 
The author also thanks Ben Jones, Georgia Karagiorgi, and Joe Grange 
for the careful reading of this manuscript.
\end{theacknowledgments}
\bibliographystyle{aipproc}   

\bibliography{sample}

\end{document}